\documentstyle[multicol,prl,aps]{revtex}
\input{epsf.tex}
\date{\today}
\begin{document}

\title{Two Gap State Density in MgB$_{2}$: A True Bulk Property or A Proximity Effect?}

\author{F. Giubileo*, D. Roditchev, W. Sacks, R. Lamy, D.X. Thanh$^{\dag}$, J. Klein\\
}
\address{
Groupe de Physique des Solides, Universites Paris 7 et Paris 6,
 UMR 75 88 au C.N.R.S.,\\ 2 Place Jussieu, 75251 Paris, France }
\address{
 *Physics Department and INFM Unit,
University of Salerno, via S. Allende, 84081 Baronissi (SA),
Italy}
\address{ $^{\dag}$Institute of Materials Science, Hanoi, Vietnam}
\author{S. Miraglia, D. Fruchart and J. Marcus$^{\ddag}$}
\address{
Laboratoire de Cristallographie, and $^{\ddag}$LEPES, CNRS, BP
166, 38042 Grenoble, France }
\author{Ph. Monod}
\address{
Laboratoire de Physique du Solide, UPR 005 au CNRS, ESPCI, 75005
Paris, France }

\maketitle

\begin{abstract}
We report on the temperature dependence of the quasiparticle
density of states (DOS) in the simple binary compound MgB$_{2}$
directly measured using scanning tunneling microscope (STM). To
achieve high quality tunneling conditions, a small crystal of
MgB$_{2}$ is used as a tip in the STM experiment. The ``sample''
is chosen to be a 2H-NbSe$_{2}$ single crystal presenting an
atomically flat surface. At low temperature the tunneling
conductance spectra show a gap at the Fermi energy followed by two
well-pronounced conductance peaks on each side. They appear at
voltages V$_{S}\simeq \pm 3.8$ mV and V$_{L}\simeq \pm 7.8$ mV.
With rising temperature both peaks disappear at the $T_{C}$ of the
bulk MgB$_{2}$, a behavior consistent with the model of two-gap
superconductivity. The explanation of the double-peak structure in
terms of a particular proximity effect is also discussed.
\end{abstract}

\vskip 2mm


\vskip 3mm


\begin{multicols}{2}

Since January of this year the scientific community has expended
considerable effort in order to understand the origin of
superconductivity in the simple binary compound MgB$_{2}$
\cite{Nagamatsu}. Its high critical temperature $T_{C}$ = 39 K is
close to or even above the upper theoretical value predicted by
the BCS theory \cite{McMillan1}. This was a strong argument to
consider MgB$_{2}$ as a non-conventional superconductor. However,
soon after the discovery of superconductivity in MgB$_{2}$, a
significant boron isotope effect has been observed \cite{Budko}.
Subsequently, a large body of experimental work has been reported
(nuclear spin-lattice relaxation rate measurements
\cite{Kotegawa}, inelastic neutron scattering measurements
\cite{Osborn}, specific heat measurements \cite{Walti,Kremer}),
all supporting an $\it{s}$-wave symmetry of the order parameter.
On the other hand, measurements of the temperature dependence of
the penetration depth \cite{Li,Panagopoulos} and of the microwave
surface resistance \cite{Zhukov} imply the possibility of an
unconventional superconductivity. Many attempts to determine the
superconducting gap have also been reported, using different
techniques: tunneling spectroscopy
\cite{Karapetrov,Rubio,Sharoni,Schmidt}, photoemission
spectroscopy \cite{Tsuda,Takahashi}, Raman measurements
\cite{Bohen}, etc. Although the results seem to indicate an
isotropic order parameter, they are controversial on the gap
width.

Adding to this controversy, the existence of two superconducting
gaps in MgB$_{2}$ was predicted theoretically by Liu {\it et al.}
\cite{Liu}. The electronic structure of  MgB$_{2}$ is quite
complex, in particular the Fermi surface presents both quasi-2D
cylindrical sheets and a 3D-tubular network, which raises the
possibility of having, in the clean limit, two distinct gaps, both
closing at the same critical temperature.

In this Letter we give a direct experimental evidence of the
existence of two distinct gaps on the surface of superconducting
MgB$_{2}$. Using our home-built low temperature STM we study the
tunneling junction obtained between one small crystal of MgB$_{2}$
mounted as the STM tip, and an atomically flat 2H-NbSe$_{2}$
single crystal. At low temperature the tunneling conductance
spectra clearly show a gap followed by a two-peak structure. The
temperature dependence of the tunneling spectra, measured with
such a superconducting MgB$_{2}$ tip, clearly demonstrates that
two different gap-features coexist up to the critical temperature
of the bulk MgB$_{2}$.

Up to now, there has been no clear experimental evidence for the
existence of two gaps in MgB$_{2}$. Characterisation of its
normal- and superconducting- states has been done by thermodynamic
measurements \cite{Wang}. As a result, the existence of a
multiple-valued gap was suggested. Photoemission spectroscopy has
also been performed on MgB$_{2}$ \cite{Tsuda}, revealing a
spectral shape that cannot be explained by the existence of a
simple isotropic gap. However, the low energy resolution, together
with the observation that both gaps were smaller than the BCS
value (5.9 mV), did not allow any firm conclusion about the
superconducting DOS of bulk MgB$_{2}$.

On the other hand, vacuum tunneling spectroscopy is a technique
allowing a direct measure of the quasiparticle DOS with high
energy resolution. Moreover, most of the studies having been done
on naturally inhomogeneous granular samples, Scanning Tunneling
Spectroscopy (STS) appears as the major tool with its high spatial
and energy resolution. Double gap structures in the differential
conductance spectra have been indeed observed by STS
\cite{Giubileo} but only locally in some selected regions of a
granular sample of MgB$_{2}$. This observation alone does not
allow one to elucidate the nature of such a feature in the
conductance spectra. While it is consistent with two-gap
superconductivity in MgB$_{2}$, it could in principle have
different origins. For example, two tunneling terms may exist, one
small, indeed reflecting the DOS of the bulk material, and another
dominant state density of the weakened superconductivity of the
surface. The existence of such a weakened layer on the very
surface was already suggested in previous studies
\cite{Karapetrov,Schmidt}. It is not excluded  that the double-gap
DOS originates from the superconductivity induced by the proximity
effect in a thin metallic surface layer \cite{McMillan2}.

The study of the temperature dependence of the tunneling spectra
is therefore necessary to clarify the origin of these particular
features. This requires an STM experiment in which the tunneling
junction remains stable, unaffected by the thermal drift due to
the temperature variations, which is almost impossible to achieve
with powder-based inhomogeneous samples. As a result of such an
inhomogeneity, small lateral movements of the tip due to the
thermal expansion of the STM unit cause dramatic changes in the
junction characteristics. The most straightforward solution to
avoid the effect of lateral drift would be the use of high quality
thin films or single crystals of MgB$_{2}$; the sample surface
being in such a case flat and the DOS spatially homogeneous. These
samples are still, unfortunately, unavailable.

As a simple measure, we inverted the tunneling junction in our
experimental setup. A small (about 50 micron size) crystal of
MgB$_{2}$ is glued with silver paint on the flat top of a
mechanically etched Pt/Ir wire, and a 2H-NbSe$_{2}$ single crystal
is taken as the sample (fig. 1 (a)). NbSe$_{2}$ is a layered
material presenting, after cleavage, an atomically flat and highly
inert surface. It is a conventional superconductor with
T$_{C}=\,7.2\,$K and thus, for higher temperatures the
experimental configuration corresponds to a S-I-N junction. At
lower temperatures it is a S-I-S one, but this fact does not
affect the spectroscopy for the temperature range studied, since
NbSe$_{2}$ is characterized by a small gap, practically smoothed
out at $T\,=\,4.2\,$K.

Geometrically such a junction is stable with respect to thermal
drift, due to a fixed tip-surface orientation. Moreover,
2H-NbSe$_{2}$ being a well-known and widely studied material, the
quality of the MgB$_{2}$ tip is easily tested by direct STM
imaging. Such an image is presented in Fig.1(b). Both atomic and
Charge Density Wave (CDW) patterns, characteristic of
2H-NbSe$_{2}$, are evident there, and even a point defect locally
perturbing the translational order of the CDW is resolved. Such a
situation clearly corresponds to vacuum tunneling. The high
quality of the junction is independently confirmed by the flat
spectral background observed in the $dI(V)/dV$ curves at
$T\,=\,4.3\,$K, as in Fig. 1(c)\cite{spectra}. These two results
unambiguously show that the STM tip apex is not contaminated and
manifests a metallic behaviour. In the spectrum of Fig. 1(c), the
double gap features are  clearly observed: well defined coherence
peaks appear at $V_{S} \simeq 3.8$ mV, followed by bumps at higher
energy $V_{L} \simeq 7.8$ mV. The spectra show very few states
inside the gap, they are independent on tunneling resistance $R_T$
and on lateral position of the tip [see inset in Fig. 1(c)]. The
curves are almost identical to those previously observed in
standard S-I-N geometry on a granular sample of MgB$_{2}$ using
Pt/Ir tip \cite{Giubileo}.

\begin{figure}[h]
 \centering
 \epsfxsize=7.5 cm \epsfbox{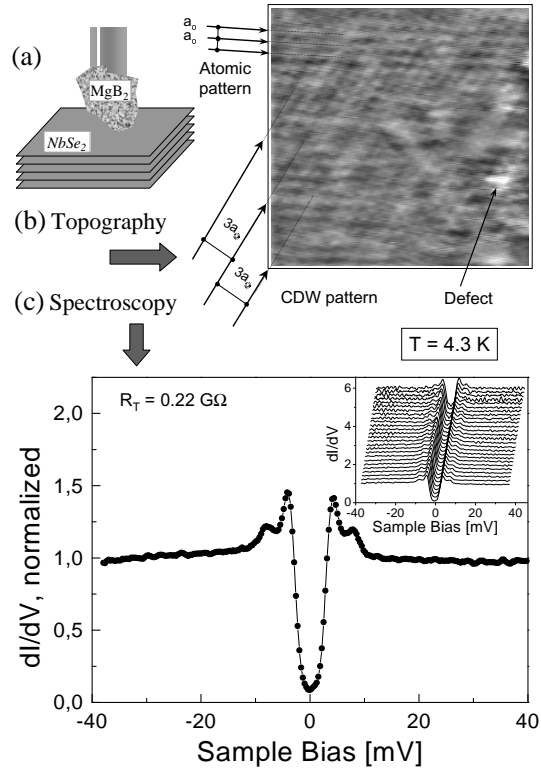}
  \caption{(a) Experimental setup.  (b) $8\,nm $ $\times$ $8\,nm $
topographic image of sample surface ($T\, = \, 4.3\, $K,  $I_T\, =
\,150\, $pA at $V_{bias}\,=-18\,$mV). The atomic lattice and a CDW
pattern are resolved. (c) Typical $dI(V)/dV$ spectrum. Inset:
series of spectra taken along a 10 nm line.} \label{fig1}
\end{figure}

Using such a high quality tunnel junction, we performed the
tunneling spectroscopy in the temperature range between 4.3 K and
45 K. The results of this experiment are reported in Fig. 2. The
overall trend is a smoothing of the gap features, and an increase
of the number of states inside the gap, with rising temperature.
The curves up to 10 K have a clear double-peak shape, confirming
that it is an intrinsic behavior of the superconducting MgB$_{2}$
electrode. For higher temperature values, the filling of the gap
continues, however. Enlarging the scale of the curves measured for
$T > 30 $K in Fig. 2, it is evident that the depletion at the
Fermi level is still present.

\begin{figure}[h]
 \centering
 \epsfxsize=7.5 cm
\epsfbox{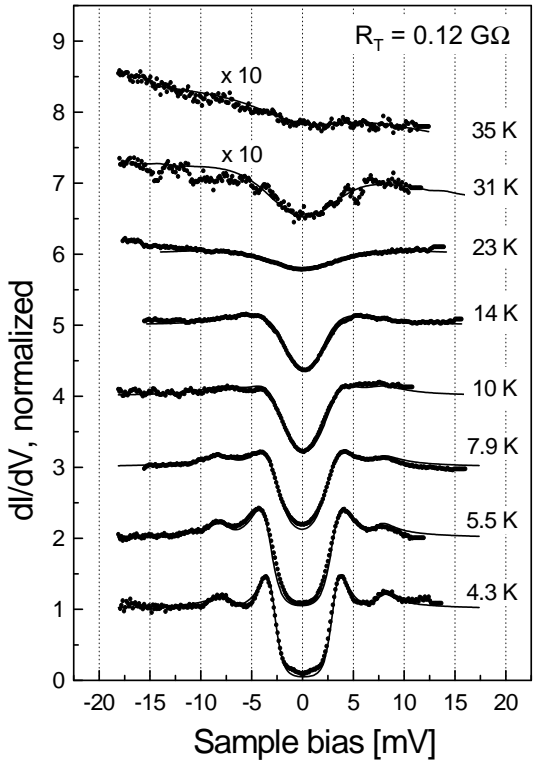}
 \caption{Temperature dependence of tunneling conductance spectra
in the range between 4.3 K and 35 K. All experimental curves
(black dots) are fitted by a sum of two weighted BCS-shape DOS
(solid lines). The spectra are shifted by unity for clarity. The
curves at T = 31 K and T = 35 K are reported magnified by a factor
10.}\label{figure2}
\end{figure}

Such a temperature dependence is non-trivial. To obtain a more
precise description, a proper fit to the tunneling spectra is very
useful. As a first step, we fit the experimental data in Fig. 2 by
considering the weighted sum of two contributions ($\sigma^{(L)}$,
$\sigma^{(S)}$) corresponding to two isotropic BCS-like DOS. The
fitting formula is written:

\begin{equation}\label{sigma}
  \sigma (V,T)=\sum_{x=L,S}C_{x}\cdot \sigma^{(x)}(V,T)
\end{equation}
where $\sigma^{(x)}$ is the tunneling conductance for the S-I-N
geometry, considering also the effect of the thermal smearing, the
damping parameter $\Gamma$, and the weight factors $C_{x}$ of the
two contributions ($C_{L}+C_{S}=1$). The fit parameters at
$T\,=\,4.3\,$K are $\Delta_{L} = 7.5$ mV, $\Delta_{S} = 3.5$ mV,
$C_{L}=0.9$ ($C_{S}=0.1$), and $\Gamma = 0.15$ mV, where $\Gamma$
is the same for both densities \cite{Thanh}. These values give a
BCS ratio ($2\Delta/k_{B}T_{C}$) of 4.5 for the large gap and 1.9
for the small one, i.e. respectively well above and well below the
weak coupling limit, 3.52.

Once determined from the fits at $T\,=\,4.3\,$K, the weight
factors ($C_{L}$ and $C_{S}$) and $\Gamma$ are kept fixed in all
other fits for higher temperatures. Thus, $\Delta_{L}$ and
$\Delta_{S}$ remain the only two free parameters. The success of
this simple approach is evident in Fig. 2, all fits (solid lines)
concur extremely well with the experimental data. The two curves
for $T > 30\,$K are enlarged by a factor ten, being near the
superconducting transition. In fact, in this temperature region,
the contribution of the small gap is strongly smeared by the
temperature, while the effect of the large gap remains small due
to its low spectral weight $C_{L}=0.1$. In spite of this, even at
these temperatures it is still possible to determine a value for
each gap although with larger uncertainty.

In comparison to the good match between the experimental data and
the fit curves, one may suppose that the spectra measured for $T >
15\,$K are not double gapped. To clarify this, we fitted the
spectra of Fig. 2 by only one BCS-like DOS. In this case one
obtains an extremely large smearing term ($\Gamma > \Delta$), in
contradiction with the low temperature results. In any case, the
latter fits are not as satisfactory as those using two BCS terms.

\begin{figure}
\centering \epsfxsize=7.0 cm
 \epsfbox{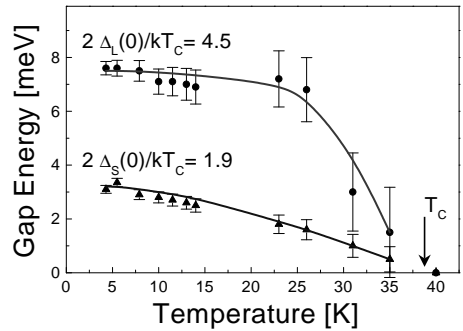}
\caption{Evolution of the gaps with the temperature. Both gaps
close near the $T_{C}$ of the bulk MgB$_{2}$.} \label{figure3}
\end{figure}

In Fig. 3, the gap widths derived from the fit values are plotted
as a function of temperature. Here the uncertainty bars are
determined independently for each data set. It is immediately
evident that both gaps vanish at a temperature close to the
$T_{C}$ ($\sim$ 39 K) of bulk MgB$_{2}$. The BCS ratio
$2\Delta(0)/k_{B}T_{C}$, extracted from the 4.3 K spectra, is 4.5
and 1.9 respectively for $\Delta_{L}$ and $\Delta_{S}$. The two
gaps vary in a different way with temperature: the small gap
reduces more rapidly than the large one. Thus, the ratio
$\Delta_{L}$/$\Delta_{S}$ between the two gaps increases with
temperature.

Recently, it has been suggested \cite{Giubileo} that, in
principle, the particular tunneling conductance spectra showing
two gap features may originate from different physical scenarios.
One possibility is the presence at the MgB$_{2}$ surface of a
contaminated layer responsible for a weak superconductivity. Of
course, this could explain the observation of double gap features
as due to two tunneling channels (one in the weak layer and
another one directly in the bulk material), but cannot account for
the small gap survival right up to the bulk $T_{C}$.

On the other hand, these data seem to confirm the existence of a
multiple-gap superconductivity, as theoretically predicted,
originating from the complicated Fermi surface of MgB$_{2}$
together with an inter-band anisotropic coupling. In this
scenario, the two gap features represent a bulk superconducting
property, giving the first direct experimental proof that
MgB$_{2}$ is the precursor of a new class of multiple-gap
superconductors. Although the possibility that the presence of two
overlapping bands can result in the observation of a two-gap
spectrum for conventional superconductors was predicted many years
ago \cite{Suhl}, until now no experimental evidence has been found
for conventional materials. This is probably because their usually
large coherence length has made the appearance of two-gap spectra
unrealistic, being difficult to realize the clean limit condition.
Only the unconventional Nb-doped SrTiO$_{3}$ has revealed a
two-gap spectrum \cite{Binnig}.

These conclusions must still be taken with a grain of salt
\cite{salt} and one must discuss a second possible interpretation
of the spectra we described here. It is generally claimed that in
the case of superconductivity induced by the proximity effect in a
metallic surface layer, the large gap feature is connected to the
bulk superconductivity, while the proximity induced small gap
should be related to a lower critical temperature. This is correct
only when the metallic layer is not ``too thin''. If the ratio of
the mean free path to layer thickness becomes large, and the N-S
interface presents a small transmitivity, it is possible that the
smaller induced gap shows a critical temperature close to or even
equal to the bulk transition temperature. This behavior has been
already observed in tunneling experiments on tin-lead sandwiches
\cite{Gilabert}. However, the fact that the width of the induced
gap depends on the metallic layer thickness, suggests that this
could not be the right scenario to explain our observations.
Indeed our data, in addition to those reported for low temperature
\cite{Giubileo}, indicate statistically a value of   3.5 mV $\pm$
0.5 mV for the small gap with no large variations, which would be
expected in the case of random thickness values of the ``normal
layer''.

Very recently it has been also suggested \cite{Manske} that the
double structure observed at low temperature in the tunneling
spectra can be explained by the existence of a low-frequency
phonon mode at 8 mV revealed in the inelastic neutron scattering
experiment \cite{Sato}. Although the calculation \cite{Manske} is
in good agreement with the low temperature spectra, the
temperature dependence reported here and the data in Fig.2 of
\cite{Giubileo} contradict this model.

In conclusion, we have reported the temperature dependence of the
tunneling DOS on superconducting MgB$_{2}$. To avoid the problem
of thermal drift and to obtain optimal tunneling conditions we
have realized an STM experiment with a superconducting tip (small
MgB$_{2}$ crystal) and a normal and atomically flat sample.
Double-gap features are clearly observed in the conductance
spectra, and their temperature dependence showed that both close
at the bulk $T_{C}$, strongly suggesting the existence of two-gap
superconductivity.

This work was supported by the Project ACI ``Nano\-structures'' of
the French Ministry of Research.

\end{multicols}

\begin{references}

\bibitem{Nagamatsu} J. Nagamatsu {\em et al.}, Nature {\bf 410}, 63
(2001)
\bibitem{McMillan1}  W.L. McMillan, Phys. Rev. {\bf 167}, 331 (1968).
\bibitem{Budko} S. L. Bud'ko {\em et al.}, Phys. Rev. Lett. {\bf 86}, 1877 (2001).
\bibitem{Kotegawa} H. Kotegawa {\em et al.}, cond-mat/0102334
\bibitem{Osborn} R. Osborn {\em et al.}, cond-mat/0103064.
\bibitem{Walti} C. Walti {\em et al.}, cond-mat/0102522.
\bibitem{Kremer} R. K. Kremer, B. J. Gibson, and K. Ahn, cond-mat/0102432.
\bibitem{Li} S. L. Li {\em et al.}, cond-mat/0103032.
\bibitem{Panagopoulos} C. Panagopoulos {\em et al.}, cond-mat/0103060.
\bibitem{Zhukov} A. Zhukov et al., cond-mat/0103587.
\bibitem{Karapetrov}  G. Karapetrov {\em et al.}, Phys. Rev. Lett. {\bf 86},
4374 (2001).
\bibitem{Rubio} G. Rubio-Bollinger, H. Suderow, and S. Vieira, Phys. Rev. Lett. {\bf 86},
5582 (2001).
\bibitem{Sharoni}  A. Sharoni, O. Millo, and I. Felner, Phys. Rev. {\bf B 63}, 220508(R) (2001).
\bibitem{Schmidt}  H. Schmidt {\em et al.},  Phys. Rev. {\bf B 63}, 220504(R) (2001).
\bibitem{Tsuda} S. Tsuda, {\em et al.}, cond-mat/0104489.
\bibitem{Takahashi}  T. Takahashi {\em et al.}, Phys. Rev. Lett. {\bf 86}, 4915 (2001).
\bibitem{Bohen} K.-P. Bohnen, R. Heid, and B. Renker, Phys. Rev. Lett. {\bf 86}, 5771 (2001).
\bibitem{Finnemore}  D.K. Finnemore {\em et al.}, Phys. Rev. Lett. {\bf 86}, 2420 (2001).
\bibitem{Liu}  A. Y. Liu, I. I. Mazin, and J. Kortus, cond-mat/0103570.
\bibitem{Wang} Y. Wang, T. Plackowski, and A. Junod, Physica C 355
(2001) 179.
\bibitem{Giubileo} F. Giubileo {\em et al.}, cond-mat/0105146.
\bibitem{McMillan2}  W.L. McMillan, Phys. Rev. {\bf 175}, 537 (1968).
\bibitem{spectra} All presented  $dI/dV$ spectra are direct derivative
of raw $I(V)$ data. No additional background substraction or
zero-bias correction was done. The spectra in Fig.1 and Fig.2 are
presented normalized by the ratio $I(V_{bias})/V_{bias}$
calculated at the sample biases $V_{bias}\,=-40\,$mV and
$V_{bias}\,=-18\,$mV respectively. Corresponding tunneling
resistances $R_T$ are indicated on each plot.
\bibitem{Thanh} The current preamplifier has an input jitter of 0.1 mV, leading to a slight
smoothing of the I(V) curves. This also implies that the intrinsic
smearing term $\Gamma$ that affects the DOS of MgB$_{2}$ in our
experiment is less then 0.05 mV.
\bibitem{Suhl}  H. Suhl, B. T. Matthias, and L. R. Walker, Phys. Rev. Lett. {\bf 3}, 552 (1959).
\bibitem{Binnig}  G. Binnig {\em et al.}, Phys. Rev. Lett. {\bf 45}, 1352 (1980).
\bibitem{salt} NaCl, ionic solid, insulator, cubic symmetry, so far
not superconducting but who knows we may all be eating a
superconductor on our French Fries.
\bibitem{Gilabert}  A. Gilabert {\em et al.}, Solid State Comm. {\bf 9}, 1295 (1971).
\bibitem{Manske} D. Manske {\em et al.}, cond-mat/0105507.
\bibitem{Sato} T. J. Sato {\em et al.}, cond-mat/0102468.

\end{references}
\end{document}